\documentclass{article}
\usepackage{epsfig}
\newcommand{\bfr}{\begin{flushright}}
\newcommand{\efr}{\end{flushright}}
 
\begin{document}
\title{Degenerate Fermions and Wilson Loop in $1+1$ Dimensions
}
\author{Kiyoshi Shiraishi\\
Institute for Nuclear Study, University of Tokyo\\
Midori-cho, Tanashi, Tokyo 188, Japan
}
\date{Canadian Journal of Physics 68 (1990) 357--360.
}
\maketitle
\begin{abstract}
We investigate the effect of finite fermion density on symmetry
breaking by Wilson loops in $(1+1)$ dimensions.  We find the
breaking and restoration of symmetry at finite density in
the models with $SU(2)$ and $SU(3)$ gauge symmetries, 
in the presence of the adjoint
fermions. The transition can occur at a finite density of
fermions, regardless of the periodic or antiperiodic boundary
condition of the fermion field; this is in contrast to the
finite-temperature ease examined by Ho and Hosotani (Nucl. Phys. B345
(1990) 445) where the boundary condition of fractional twist is
essential to the occurrence of the phase transition.
\end{abstract}

\section{Introduction}
The physics of low-dimensional systems is currently studied
by many authors. Some recent motivations for the study of
low-dimensional physics come from the investigation of string
theories \cite{1}, i.e., field theory on two-dimensional world
sheets, and conformal field theories \cite{2}.

More recently, three-dimensional systems have attracted
much interest. There are interesting subjects to study, for
example, high-$T_c$ superconductors \cite{3}, fractional statistics
\cite{4}, boson-fermion transmutation \cite{5}, topological field
theories \cite{6}, and exactly soluble gravity and the relation to the
mathematics of ``knots'' \cite{7}. In these theories, the peculiarity
in low-dimensional space-time is used fully. Note that it seems that
some gauge symmetries and ``effective'' gauge fields play crucial
roles.

On the other hand, as is stated in many cases, it is helpful
to study models with reduced numbers of degrees of freedom, which
depend on the dimensionality of the system. The solution of
the models might shed on the more compliecated problem in higher
dimensions.

In $(1+1)$ dimensional pure Yang-Mills theory there is no
dynamical degree of freedom. There is, however, a nontrivial
topological quantity if the background space is a circle. Recently,
pure Yang-Mills theory applied to a circle has been investigated by
Rajeev \cite{8}. The gauge covariant variable is often called a Wilson
loop. It is known that the Wilson loop couples to matter fields and
affects their dynamics \cite{9}.

Conversely, quantum effects of matter fields determine the
vacuum expectation value of the Wilson loop. Symmetry
breaking by Wilson loops, not only on a torus space
\cite{10,11,12,13,14,15,16,17,18,19,20,21,22} but also on a nonsimply
connected lens space \cite{23,24,25,26,27,28}, has been considered
and used in unified models of gauge theories in higher (more than four)
dimensions. Other approaches to the mechanism in string theories in
novel ways are found in refs.~\cite{29,30,31,32}.

Recently, Ho and Hosotani \cite{19} examined two-dimensional
models with $SU(3)$ gauge symmetry and fermions to investigate phase
transitions at finite temperature.

In this paper we consider nonzero fermion density in similar
systems. We show the realization of other symmetries of vacuum when
fermions are strongly degenerated. We consider Dirac fermions in
this paper. Furthermore, we deal with left-right symmetric models;
thus, anomalies have nothing to do with the phenomenon that
takes place in our models.

\section{The model and calculation}

The action that We consider in this paper is given by
\begin{equation}
I=\int dt \int_0^L dx \left(-\frac{1}{4}{\rm Tr~}F_{\mu\nu}
F^{\mu\nu}+{\rm Tr~}\bar{\psi}i\gamma^\mu D_\mu\psi\right)\,,
\label{eq1}
\end{equation}
where $L$ is circumference of the spatial circle, the field strength
is defined as, for example, $F_{01}=\partial_0 A_1-\partial_1
A_0+ig[A_0, A_1]$, and the covariant derivative is, for example,
$D_1\psi=\partial_1\psi+ig[A_1, \psi]$. Here $g$ is a gauge coupling
constant. The massless fermion field $\psi$ is in the adjoint
representation.

It has been found that the quantum effect of fermions in
the adjoint representation or the ``faithful, single-valued
representation'' of the adjoint group \cite{16} can naturally bring
about gauge-symmetry breakdown in higher dimensions. On these
occasions, the boundary condition on the field is also important in
the evaluation of quantum effects. We take the boundary conditions as
\begin{eqnarray}
A_\mu(t, x+L)&=&A_\mu(t, x)\,,\\
\psi(t, x+L)&=& e^{iB}\psi(x, L)\,,
\label{eq2}
\end{eqnarray}
where the notation is the same as in ref.~\cite{19} except for the
``twist'' parameter $B$ ($-\pi\le B\le\pi$). The freedom of
the global group transformation has been rotated away in (\ref{eq2}).

The physical quantity is the Wilson loop
\begin{equation}
W=P\exp\left(ig\int_0^L dx\, A_1\right)\,,
\end{equation}
where $P$ means the path-ordered product. In vacuum, we suppose that
the vacuum expectation value $\langle F_{01}\rangle=0$. Nevertheless,
$\langle W\rangle$ can be nontrivial.

In two dimensions, there is no dynamical gauge boson; thus,
hereafter we call the phenomenon of $\langle W\rangle$ acquiring a
nontrivial value as ``symmetry breaking.'' Moreover, we define
symmetry of the vacuum by the subgroup of elements that commute with
$\langle W\rangle$.

In the following, we consider nonzero fermion density. For this
purpose, we calctulate the thermodynamic potential $\Omega$ for the
system at one-loop level. It is well known that the
thermodynamic potential is expressed formally as \cite{33,34}
\begin{equation}
\Omega=-\frac{1}{\beta}\ln Z\,
\end{equation}
with
\begin{equation}
Z={\rm tr~} \{\exp[-\beta(\hat{H}-\mu\hat{N})]\}\,,
\end{equation}
where $\hat{H}$ is a Hamiltonian operator and $\hat{N}$ is a
number operator in the theory. The trace is to be taken on the
Hilbert space of states of the theory under consideration.  $\beta$
is the inverse temperature, and $\mu$ is the chemical potential
associated with $N$. In the present model $N$ is taken as the
fermion number;
\begin{equation}
\hat{N}\sim {\rm Tr~} \int
dx\,\psi^\dagger\psi\,.
\end{equation}
  
In the field theoretical description at the one-loop level, $\Omega$
is formally written in our model as
\begin{equation}
\Omega=-\frac{1}{2}{\rm tr~}\ln (i\gamma^\nu D_\nu)\,,
\end{equation}
where $D_0=\partial_0+ig[\langle A_0\rangle,~]$ and
$D_1=\partial_0+ig[\langle A_1\rangle,~]$ with
$\langle A_0\rangle=-i\mu$ \cite{33}.
${\rm tr}$ means sum over the oscillator degrees of freedom and the
trace over the gamma matrix at the same time. In two dimensions,
the contribution of the gauge field to the potential is cancelled by
the contribution of the associated ghost fields.

First we consider $SU(2)$ gauge symmetry. By using gauge
transformation, we can parametrize a diagonal form of the vacuum
gauge field:
\begin{equation}
\langle A_1\rangle=\frac{1}{gL}
\left(
\begin{array}{rr}
\theta & 0 \\
0 & -\theta
\end{array}\right)\,.
\end{equation}

The termodynamic potential or the effective potential in
terms of $\theta$ with $\mu\ne 0$ in this background is given by
\cite{33,34} 
\begin{eqnarray}
\Omega/L&=&\frac{2}{\pi
L^2}\sum_{l=1}^\infty\frac{1}{l^2}\{\cos[l(2\theta-B)]+
\cos[l(2\theta+B)]+1\}\nonumber \\
& &-\frac{1}{\beta L}\sum_{-\infty}^\infty\left\{
\ln\left[\prod_{i=1}^3(1+e^{\beta(\mu-M_i)})(1+e^{-\beta(\mu+M_i)})
\right]\right\}\,,
\label{eq9}
\end{eqnarray}
where $M_1=|2\pi l+2\theta-B|/L$, $M_2=|2\pi l-2\theta-B|/L$ and
$M_3=|2\pi l|/L$.  The expression of the vacuum energy in the
first part of (\ref{eq9}) is obtained after the regularization in the
standard way \cite{34,13}.
Obviously the potential in the region $2\pi\le 2\theta\le 4\pi$
has the same form as in the region $0\le 2\theta\le 2\pi$ because
of $Z_2$ symmetry \cite{9}.  Therefore we restrict ourselves to the
examination of the region $0\le 2\theta\le 2\pi$.

At $T=\beta^{-1}=0$ and $\mu=0$, performing the sum over $l$,
we can obtain the vacuum energy as
\begin{equation}
V_0=\left(\frac{\Omega}{L}\right)_{\mu=0}=-
\left\{
\begin{array}{ll}
~[(2\theta)^2+(\pi-B)^2]/(\pi L^2) & 0\le 2\theta\le B \\ 
~[(2\theta-\pi)^2+B^2]/(\pi L^2) & B\le 2\theta\le 2\pi-B \\ 
~[(2\theta-2\pi)^2+(\pi-B)^2]/(\pi L^2) & 2\pi-B\le 2\theta\le 2\pi
\end{array}
\right.\,.
\label{eq10}
\end{equation}

The finite density effect becomes important at low temperature.
In the limit $T\rightarrow 0$ ($\beta\rightarrow\infty$),
eq.~(\ref{eq10}) becomes \cite{13,34}
\begin{equation}
\frac{\Omega}{L}=V_0-\frac{1}{L}\sum_{i=1}^3\sum_{M_i\le\mu}
(\mu-M_i)\,.
\label{eq11}
\end{equation}
The sum over $l$ is taken only in the finite range where $M_i\le\mu$
is satisfied. The quatization of the fermion number is easily read
from the expression. (Note: The fermion number is given by
$-\partial\Omega/\partial\mu$.)

Here, we pay attention to the case that $B=0$, i.e., the periodic
boundary condition on the fermion field. At high temperatures, no
phase transition occurs, as shown by Ho and Hosotani \cite{19}.
The minimum of the potential in terms of $\theta$ is located at
$2\theta=\pi$ at any finite temperature if $B=0$.  In our case with
degenerate fermions, a different aspect comes about.

The shape of the potential $\Omega/L$ is displayed in Fig.~1. The
potential is continuous but not smooth everywhere. This feature is
found only in two dimensions;
in higher dimensions, the potential is smooth everywhere \cite{13}.

\begin{figure}[ht]
\begin{center}
\includegraphics[width=3.5cm]{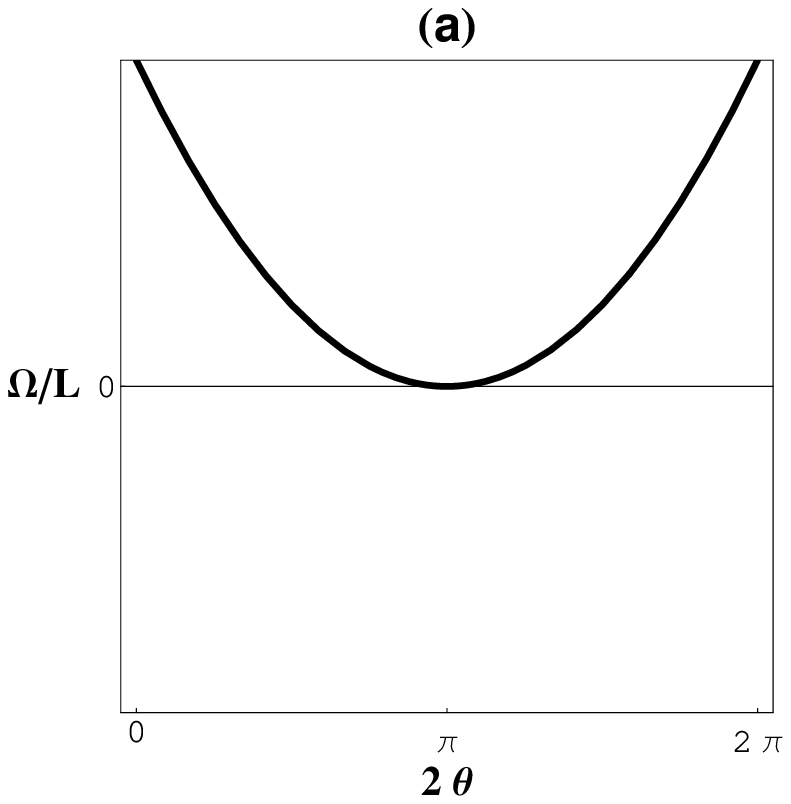}
\includegraphics[width=3.5cm]{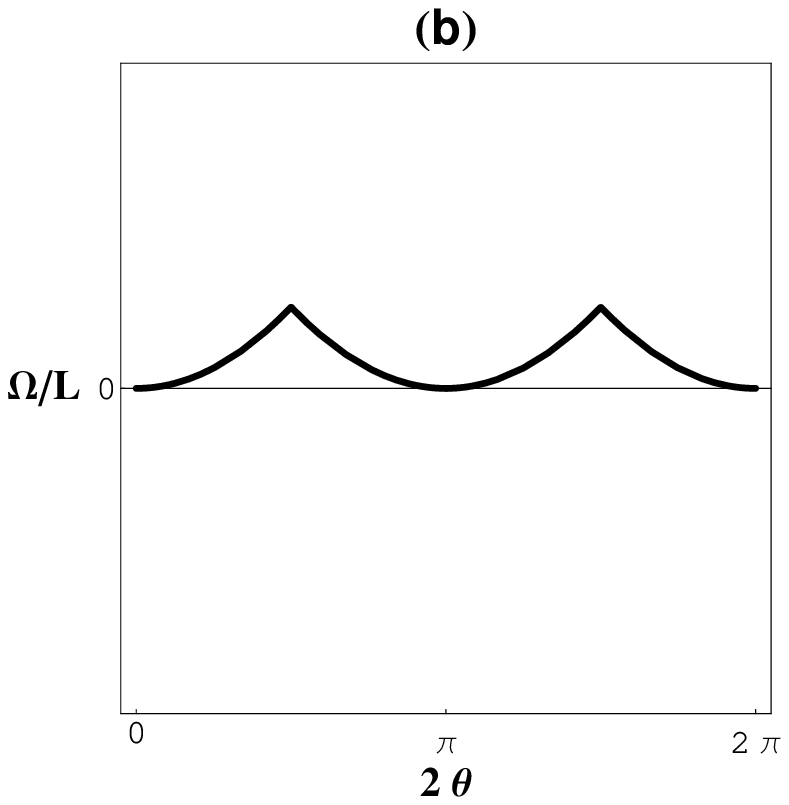}
\includegraphics[width=3.5cm]{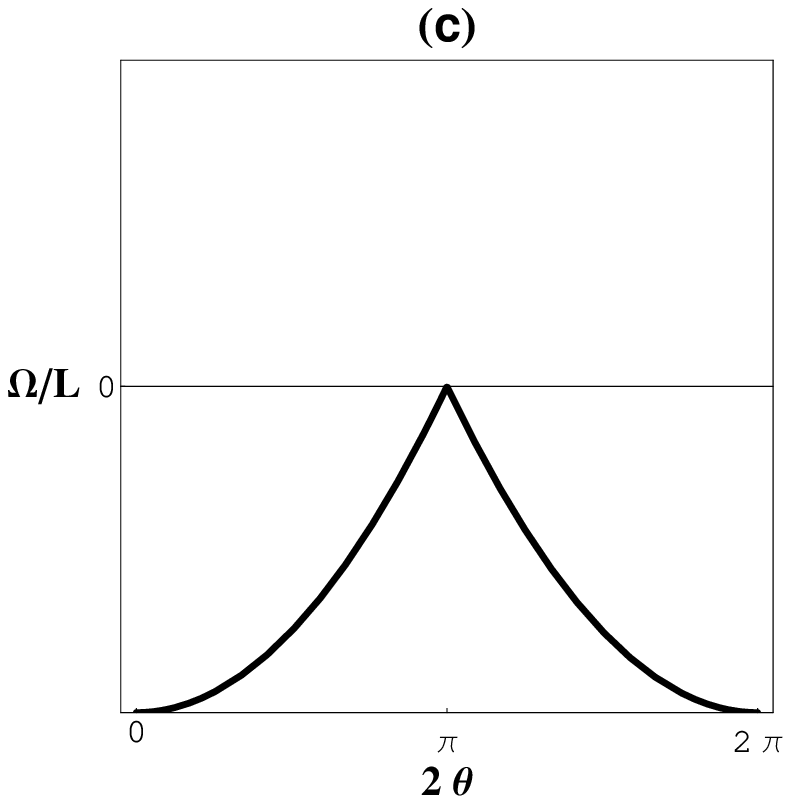}
\caption{Schematic view of the thermodynamic potentials at zero
temperature including vacuum energy as a function of $2\theta$ for
various values of the chemical potential. (a) $\mu L=0$. (b) $\mu
L=\pi/2$. (c) $\mu L=\pi$.}
\label{f1}
\end{center}
\end{figure}

Let us consider the difference in $\Omega/L$ at $2\theta=0$ and
$2\theta=\pi$. It turns out, from eq.~(\ref{eq10}) and
eq.~(\ref{eq11}), that
\begin{eqnarray}
(\Omega/L)(2\theta=0)-(\Omega/L)(2\theta=\pi)=-
\left\{
\begin{array}{ll}
(\pi-2\mu L)/L^2 & 0\le\mu L\le\pi\\ 
(-3\pi+2\mu L)/L^2 & \pi\le\mu L\le 2\pi
\end{array}
\right.\,.
\end{eqnarray}
Actually, this is a periodic function of $\mu L$.
Thus, if $1/4<(\mu L/2\pi)-[\mu L/2\pi]<3/4$ (where $[~]$ is Gauss'
symbol), the point $2\theta=\pi$ is favoured rather than $2\theta=0$.
This is a new vacuum at finite fermion density.

Next we consider a model with $SU(3)$ symmetry. The
form of the action is the same as eq.~\ref{eq1}.
The vacuum  gauge field is parametrized as \cite{19}
\begin{equation}
\langle A_1\rangle=\frac{1}{gL}
\left(
\begin{array}{rrr}
\theta_1 &  &\\
  & \theta_2 & \\
 & & \theta_3
\end{array}\right)\,,
\end{equation}
with $\theta_1+\theta_2+\theta_3=0$.

At zero temperature and with vanishing chemical potential, the
minima of the effecive potential determine the following
symmetry of the vacuum:
\begin{equation}
\begin{array}{ll}
|B| < \pi/3, &  U(1) \times U(1)\\
\pi/3 < |B| <\pi/2, &  SU(2) \times U(1)\\
\pi/2 < |B| < \pi, & SU(3)\,.
\end{array}
\end{equation}

In the background of the strongly degenerate fermion at
zero temperature, one can find the symmetries of the vacua at
$B=0$ as
\begin{equation}
\begin{array}{ll}
|\mu L| < \pi/3, &  U(1) \times U(1)\\
\pi/3 < |\mu L| <\pi/2, &  SU(2) \times U(1)\\
\pi/2 < |\mu L| < \pi, & SU(3)\,.
\end{array}
\end{equation}

For general values of $B$ and $\mu L$, the symmetry is exhibited
in Fig.~2.

\begin{figure}[ht]
\begin{center}
\includegraphics[width=5
cm]{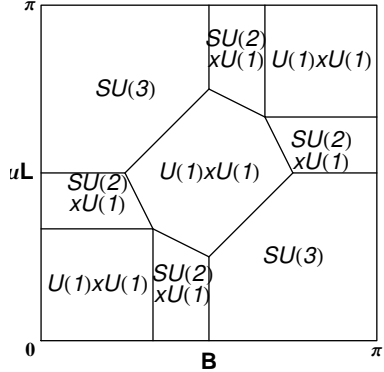}
\caption{Phase structure of $1+1$ dimensional $SU(3)$ model
with ``twisted'' adjoint fermions at zero temperature and
nonvanishing chemical potential $\mu$.  The other region of
parameters $B$ and $\mu L$ can be obtained by taking mirror images of
this diagram.}
\label{f2}
\end{center}
\end{figure}

In the more general situations that appear in physical problems, we
must impose conservation of the fermion number. In other
words, $\mu$ is to be determined by $N$. However, we do not pursue
further analysis on complicated systems in this paper.

\section{Concluding remarks}

We wish to emphasize that a new phase exists at finite density
even if the boundary condition on the fermion is periodic or
antiperiodic. High-temperature phase transitions investigated by Ho
and Hosotani \cite{19} are characteristic of the model with matter
fields that obey the boundary condition of an intermediate twist
angle, i.e., $B \ne 0$ or $\pi$. However, the fractional twist is
rather unnatural, especially if we consider higher dimensional
theories where we use the symmetry-breaking mechanism of Wilson
loops on compact spaces. In the compactified theory, the matter
field seems massive even if $\langle W\rangle=1$. Of course, our
``phase transition'' scenario at finite density can work at unnaturally
high densities whose scales are comparable to the Planck scale if we
utilize the mechanism in unified theories.

On the other hand, from the low-dimensional field theoretical point
of view, the Wilson loop might give an interesting perspective. The
boundary conditions of the fields are effectively determined
dynamically in some sense. The finite-volume effect on the system
may equally be significant in other two- or three-(space-time)
dimensional systems. Further non-trivial connections between gauge
fields and matter fields will be clarified in the procedure presented
in this paper.


\section*{Acknowledgements}
The author would like to thank A. Nakamula for some comments.
This work is supported in part by a Grant-in-Aid for the Encouragement
of Young Scientists from the Ministry of Education, Science, and
Culture (No.~63~790~150). The author is grateful to the Japan Society
for the Promotion of Science for the fellowship. He also thanks
Iwanami F\=ujukai for financial aid.


\end{document}